\documentclass[3p,times,twocolumn]{elsarticle}

\usepackage{ecrc}

\volume{00}

\firstpage{1}

\journalname{Nuclear and Particle Physics Proceedings}

\runauth{K. Kauder for the STAR Collaboration}

\jid{nppp}

\jnltitlelogo{Nuclear and Particle Physics Proceedings}

\usepackage{amssymb}
\usepackage{amsmath}

\usepackage{lineno}

\usepackage[figuresright]{rotating}

\begin{document}

\begin{frontmatter}



\dochead{}

\title{Measurement of the Shared Momentum Fraction $z_g$ using Jet Reconstruction in p+p and Au+Au Collisions with STAR}


%\renewcommand{\floatpagefraction}{0.9}

\author{K. Kauder for the STAR Collaboration}

\address{Wayne State University, Detroit, MI-48201, USA}

\begin{abstract}
One key difference in current energy loss models lies in the treatment of the Altarelli-Parisi, AP, splitting functions.
It has been shown that the \emph{shared momentum fraction}, henceforth called \emph{Jet Splitting Function} $z_g$
 as determined by the \mbox{SoftDrop} grooming process 
can be made a Sudakov-safe measurement of the symmetrized AP functions in p+p collisions. 
The STAR collaboration presents the first $z_g$ measurements at $\sqrt{s_{NN}}=200$~GeV in p+p and Au+Au collisions,
where in Au+Au we use the specific di-jet selection introduced in our previous momentum imbalance measurement. 
For a jet resolution parameter of $R=0.4$, these di-jet pairs were found to be significantly imbalanced with respect to p+p,
yet regained balance when all soft constituents were included.
We find that within uncertainties there are no signs of a modified Jet Splitting Function on trigger or recoil side of this di-jet selection.

\end{abstract}


\end{frontmatter}



\section{Introduction}
\label{sec:introduction}

Jet reconstruction algorithms and techniques used to correct for the underlying event have been primarily
developed by the particle physics community as a robust tool to access parton kinematics from measured final-state hadrons.
Modern approaches to extract information from the jet sub-structure pioneered
by particle physics applications have recently found their way into 
the heavy-ion field, where the dramatically larger underlying event poses unique challenges.
For an excellent review of the now ubiquitous class of infra-red and collinear safe sequential clustering algorithms
($k_T$, anti-$k_T$, Cambridge/Aachen(C/A)) and of the concepts used in this analysis, 
please refer to M. Cacciari's presentation at this conference.%\cite{cacciari_theseprocs} 

Here, the considered observable is the groomed momentum fraction $z_g$, or \emph{Jet Splitting Function},
that allows a direct measurement of a fundamental building block of pQCD in p+p collisions, 
the (symmetrized) Altarelli-Parisi splitting functions.
It emerges as a ``by-product'' of the \mbox{SoftDrop}~\cite{Larkoski:2014wba} grooming technique
used to remove soft wide-angle radiation from a sequentially clustered jet.
This is achieved by recursively declustering the jet's branching history and discarding subjets until the 
transverse momenta $p_{T,1}, p_{T,2}$ of the 
current pair of subjets fulfill the \mbox{SoftDrop} condition:
\begin{equation}
  \label{eq:SD}
  \frac{\min(p_{T,1},p_{T,2}) }{ p_{T,1}+p_{T,2}} > z_{\text{cut}}\theta^\beta,
\end{equation}
where $\theta$ is an additional measure of the relative distance between the two sub-jets.
The current analysis disregards $\theta$ by setting $\beta=0$, and 
we follow the authors' default choice $z_{\text{cut}}=0.1$.
It was shown  that for such a choice, and for a C/A clustering, 
the distribution of the resulting groomed momentum fraction, or \emph{Jet Splitting Function} 
\begin{equation}
  \label{eq:zg}
  z_g \equiv \frac{\min(p_{T,1},p_{T,2}) }{ p_{T,1}+p_{T,2}}
\end{equation}
converges to the vacuum AP splitting functions for $z>z_{\text{cut}}$ in a 
``Sudakov-safe'' manner~\cite{Larkoski:2015lea},
i.~e. independent of non-perturbative physics in the UV limit and eliminating the $\mathcal{O}(\alpha_s)$ order.
In A+A collisions, modification of the splitting is a characteristic aspect in some classes of 
energy loss models, and the measurement of $z_g$ presented here gives qualitatively new constraints
for theoretical treatment. Alternatively, quenching of the sub-jets after a vacuum-like split could also
lead to $z_g$ modification.

\begin{figure}[h]
  \centering
  \includegraphics[width=.44\textwidth]{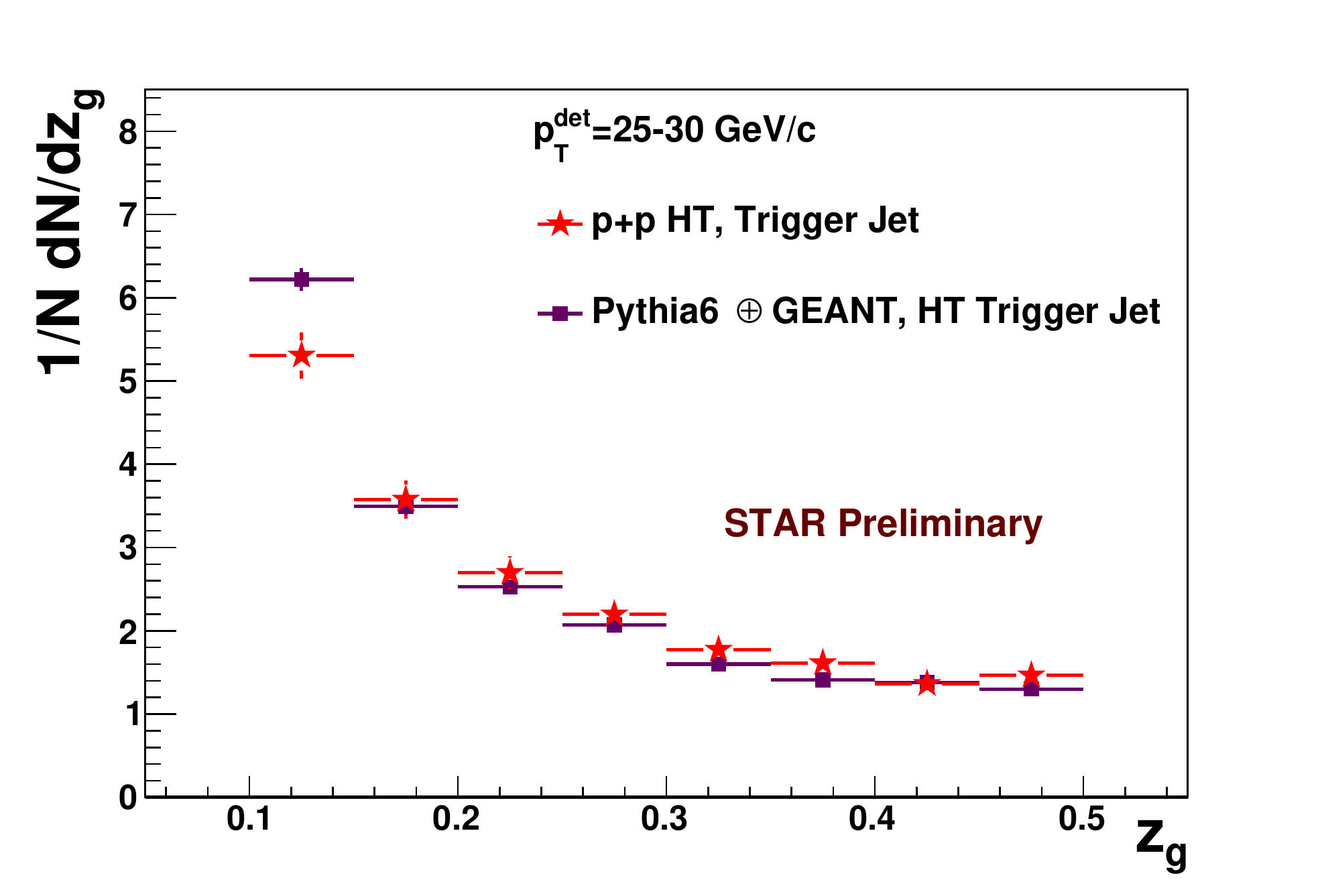}
  \caption{Trigger jets for p+p HT compared to PYTHIA predictions 
    after simulation of the STAR detector.
    Independently binned in one example $p_T^{\text{det}}$ bin at detector level.
    Error bars are statistical only.}
  \label{fig:ppraw}
\end{figure}
 
All jets are found using the anti-$k_{T}$ algorithm from the FastJet package~\cite{fastjetArea,fastjet} 
with resolution parameter $R = 0.4$.
Data selection and detector setup is identical to Ref.~\cite{Adamczyk:2016fqm}.
The data were collected by the STAR detector in p+p and Au+Au collisions at $\sqrt{s_{NN}}$ = 200~GeV in 2006 and 2007, respectively.
Constituents in the jet finding charged tracks were reconstructed with the Time Projection Chamber (TPC)~\cite{Anderson:2003ur}, 
and neutral hadrons with transverse energy $E_T$ were measured in the Barrel Electromagnetic Calorimeter (BEMC) ~\cite{Beddo:2002zx},
with a so-called full hadronic correction scheme in which the transverse momentum of any charged track that extrapolates to a tower
is subtracted from the transverse energy of that tower.
Tower energies are not allowed to become negative via this correction. 
An online High Tower (HT) trigger required $E_T > 5.4$~GeV in at least one BEMC tower.

\section{Pythia Study}
\label{sec:pythia}
A p+p simulation at $\sqrt{s}$=200 GeV  of leading jet $z_g$  was conducted using PYTHIA 6.410~\cite{Sjostrand:2006za}
with CTEQ5L pdfs~\cite{Lai:1999wy}
and PYTHIA 8.219~\cite{Sjostrand:2007gs}
with default settings. 
As an additional difference, the PYTHIA8 sample only contains stable particles in the final state while the PYTHIA6 sample also comprises
short-lived and long-lived particles since the final decay happens at a later stage in the simulation of the STAR detector.
Despite the differences, both lead to nearly identical $z_g$ distributions (not shown) and qualitatively
good agreement with the analytical solution.

\begin{figure}
  \centering
  \includegraphics[width=.44\textwidth]{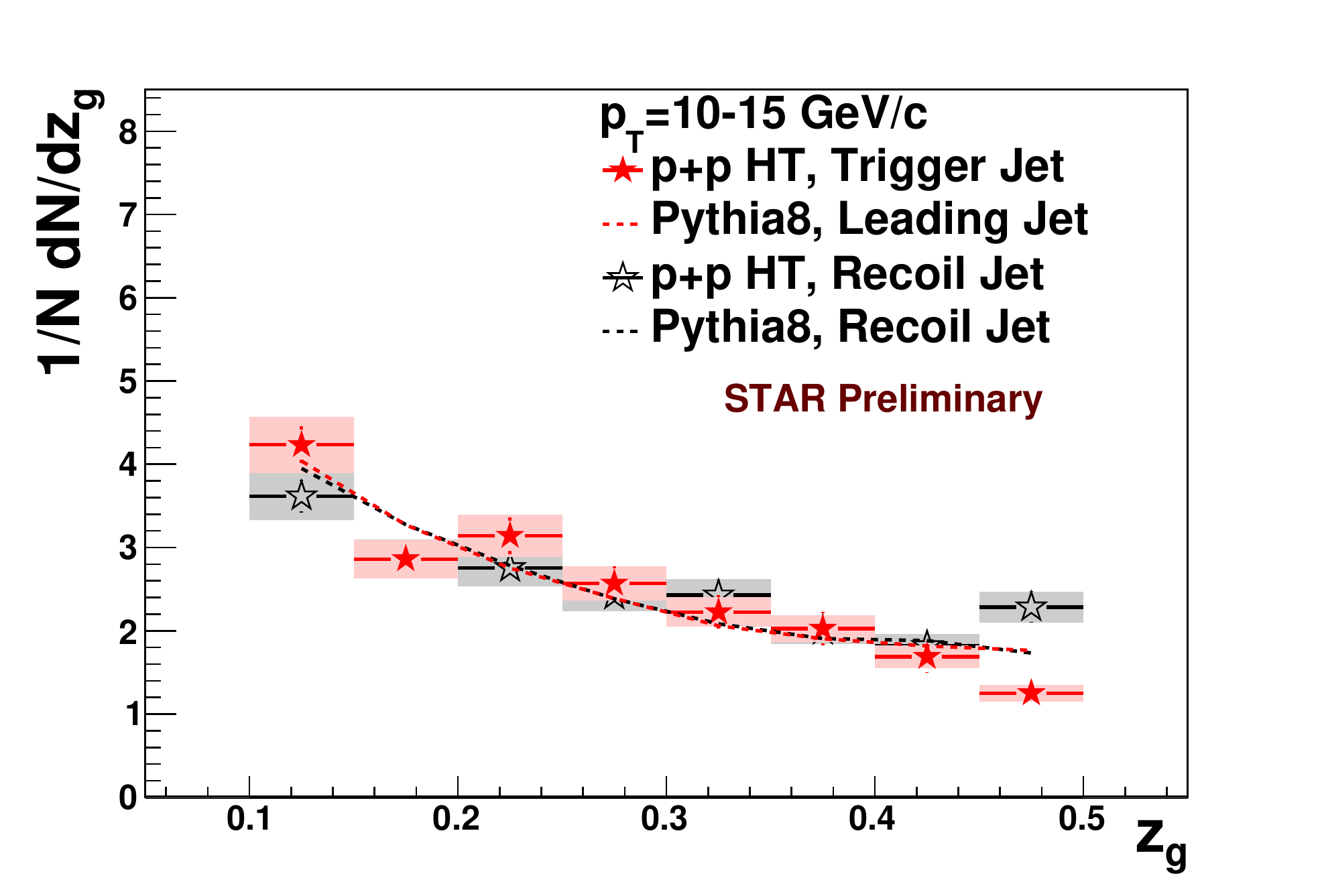}
  \includegraphics[width=.44\textwidth]{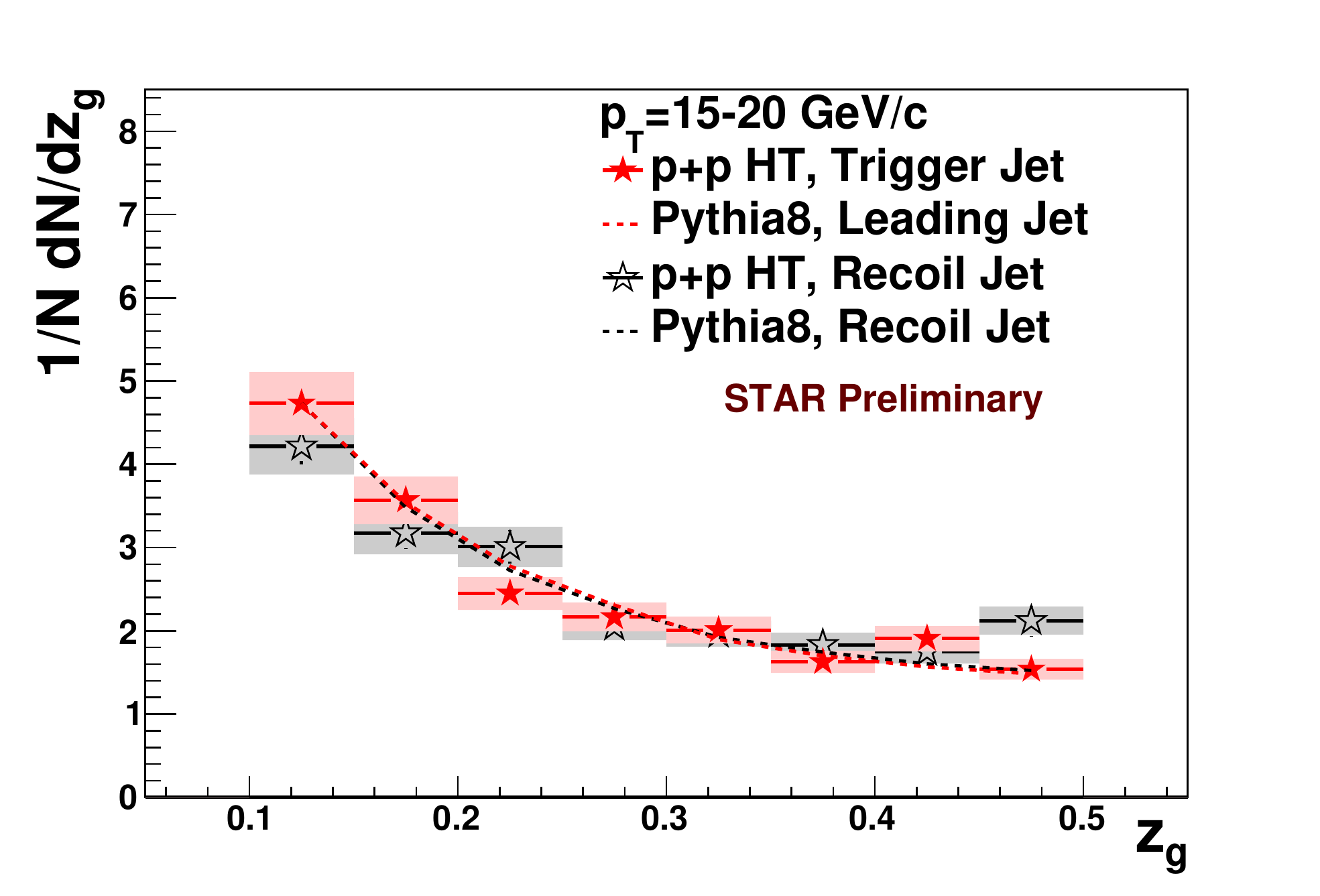}
  \includegraphics[width=.44\textwidth]{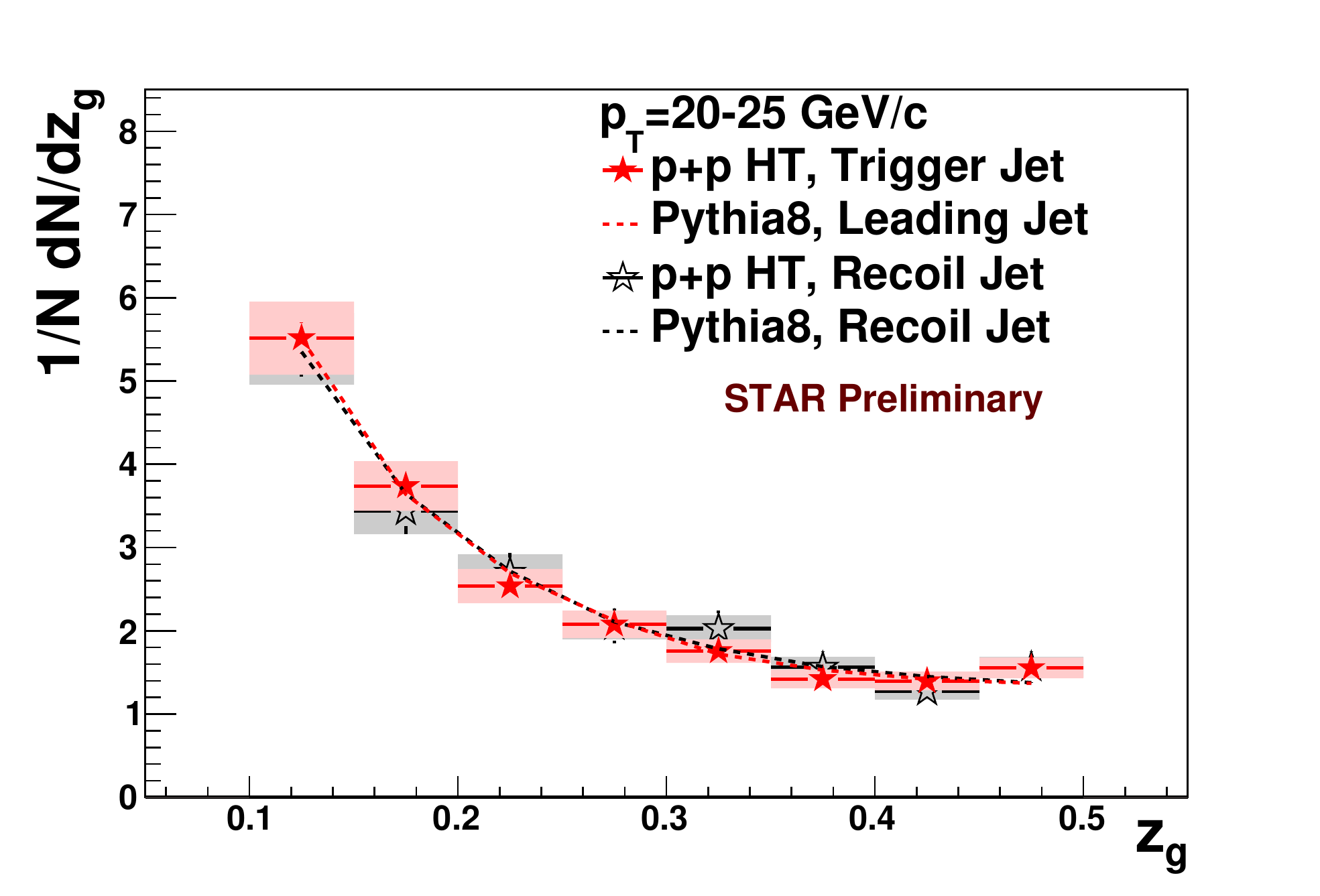}
  \includegraphics[width=.44\textwidth]{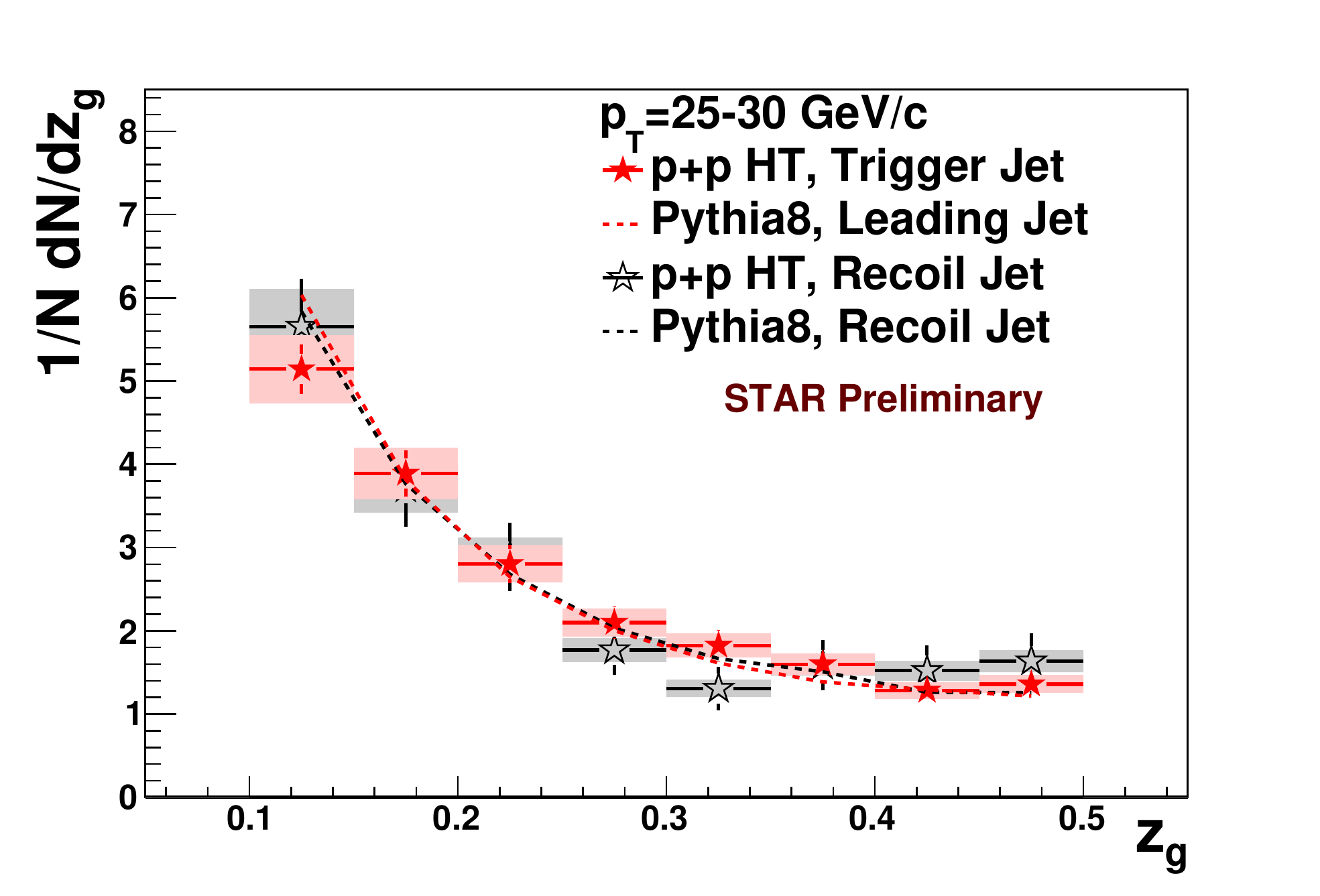}
  \caption{Corrected $z_g$ distributions for trigger (filled symbols) and recoil (open symbols) jets in p+p HT compared to PYTHIA8 (dashed lines),
    independently binned in $p_T^{\text{part}}$ bins.
    Shaded bands indicate systematic uncertainty estimate due to the jet energy scale.}
  \label{fig:ppcorr}
\end{figure}

%DiJets!!
\begin{figure*}[htbp]
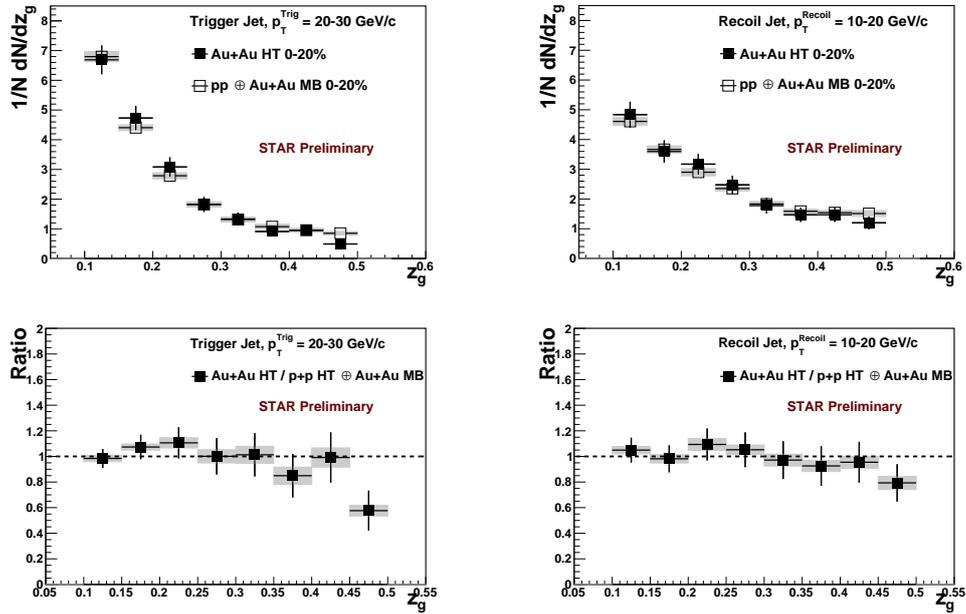

  \centering
  \includegraphics[width=.38\textwidth]{{{Groom_AuAu_HT54_HTled_Tow0_Eff0_Groom_Aj_HT54_HTled_ppInAuAuAj.DijetLeadZg2030}}}\qquad
  \includegraphics[width=.38\textwidth]{{{Groom_AuAu_HT54_HTled_Tow0_Eff0_Groom_Aj_HT54_HTled_ppInAuAuAj.DijetSubLeadZg1020}}}
  \includegraphics[width=.38\textwidth]{{{Groom_AuAu_HT54_HTled_Tow0_Eff0_Groom_Aj_HT54_HTled_ppInAuAuAj.RatioLeadZg2030}}}\qquad
  \includegraphics[width=.38\textwidth]{{{Groom_AuAu_HT54_HTled_Tow0_Eff0_Groom_Aj_HT54_HTled_ppInAuAuAj.RatioSubLeadZg1020}}}
  \caption{Top: Distributions of $z_g$ of trigger and recoil jets for Au+Au HT data (filled symbols) and p+p HT $\otimes$ Au+Au MB
    (open symbols), independently binned in $p_T^{\text{part}}$ bin.
    Bottom: Ratio between Au+Au HT data and p+p HT $\otimes$ Au+Au MB.
    Shaded bands indicate systematic uncertainty estimate due to the jet energy scale.}
  \label{fig:auau}
\end{figure*}

\section{Measurement in p+p HT}
\label{sec:pp}
To estimate the effect of a High Tower trigger in p+p, the PYTHIA8 simulation was first repeated with the additional requirement of 
a neutral 5.4 GeV/$c$ particle in the trigger jet. As expected, we found no difference on the recoil side between triggered and untriggered events. 
The trigger bias in the $z_g$ distributions disappears around $p_T=20-25$~GeV/$c$.
In this analysis, we distinguish between ``trigger'' and ``recoil'' jets 
depending on which jet contains the High Tower that fulfilled the trigger requirement. 

An example comparison at the detector level (without efficiency or smearing corrections; $p_T^{\text{det}}$)
of trigger  jets between measured p+p HT and 
the above-mentioned PYTHIA6 data after detector simulation is shown in Fig~\ref{fig:ppraw}.
For both trigger and recoil, and for all shown $p_T^{\text{det}}$ bins between 10 and 30 GeV/$c$,
we observe excellent agreement between the measured data and PYTHIA6 
when folded by the STAR detector simulation.
 
It is therefore appropriate to use a bin-by-bin correction as a first approach 
to correct for detector effects and the HT trigger bias. The corrected distributions 
are shown in Fig.~\ref{fig:ppcorr} in $p_T^{\text{part}}$ bins,
where $p_T^{\text{part}}$ refers to the value corrected to particle level.
Measurements above 30 GeV/$c$ only have reasonable statistics for trigger jets, and hence are omitted here.
The overlaid dashed lines demonstrate the $z_g$ agreement with PYTHIA8 on both trigger and recoil side for jets in p+p.
The shaded bands in Fig.~\ref{fig:ppcorr} represent the uncertainty 
due to the overall jet energy scale uncertainty of 4\%~\cite{Abelev:2007vt}. 
Note that this scale uncertainty when applied to subjets cancels out in the 
calculation of $z_g$, hence we only consider $p_{T}^{\text{part}}$ bin migration.
Nevertheless, especially at lower jet $p_T$ the presence of a High Tower leads 
to a significantly different neutral energy fraction in the trigger jet and thus in 
one of its subjets.
An evaluation of the effect of tracking efficiency and tower scale uncertainty on individual subjets and their potential (anti-)correlation 
is underway.

\section{Triggered Di+jets in Au+Au}
\label{sec:auau}

For the first $z_g$ measurement in 0-20\% central Au+Au collisions,
we focus on a di-jet selection very similar to previous $A_J$ measurements~\cite{Adamczyk:2016fqm}.
The initial definition of the di-jet pair considers only tracks and towers with $p_{T}^{\text{Cut}} >2$~GeV/$c$ in the jet reconstruction.
Due to the symmetry of a di-jet imbalance measurement, it was previously unnecessary to keep track of the High Tower.
As noted above, in this analysis, we consider the two sides of the di-jet pair separately and thus differentiate between trigger and recoil jets.
Di-jets were accepted for trigger jets with $p_T^{\text{Trig}}>20$~GeV/$c$ and recoil jets with $p_T^{\text{Recoil}}>10$~GeV/$c$;
a requirement for the trigger jet to be the leading jet was not enforced.
Kinematic cuts are made on $p_T^{\text{Trig,Recoil}}$, i.e. only considering the ``hard core'' above 2~GeV/$c$.
This constituent $p_t$ bias was relaxed in the $z_g$ calculation by using
geometrically matched (axes within \mbox{$\Delta R = \sqrt{\Delta \phi^{2} + \Delta \eta^{2} }<R$}) 
di-jet pairs reconstructed with $p_{T}^{\text{Cut}} >0.2$~GeV/$c$.
Area-based background subtraction on the matched jets was carried out during the SoftDrop algorithm
following the standard FastJet procedure~\cite{fastjet}, where the event-by-event background energy density $\rho$ is determined with
the $k_T$ algorithm with the same $R$ as the median of $p_T^{\text{jet,rec}} / A^{\text{jet}}$ of all but the two leading jets,
and the jet area $A^{\text{jet}}$ is found using active ghost particles. 

Analogous to the $A_J$ analysis, a reference data set is constructed by embedding p+p HT events into
minimum bias Au+Au events in the same centrality class (p+p HT $\otimes$ Au+Au MB). 
Thus, jets are compared with similar initial parton energies in Au+Au and p+p, and the remaining effect of background fluctuations are accounted for.
The jet energies are not corrected back to the original parton energies.
During embedding, the differences between Au+Au and p+p 
in tracking efficiency in the TPC ($90\% \pm 7\%$),
relative tower efficiency ($98\% \pm 2\%$, negligible),
and the relative tower energy scale ($100\% \pm 2\%$) are applied.
Systematic uncertainty on $z_g$ was assessed in this process by varying the relative efficiency and tower scale within their uncertainties
and is shown in the p+p HT $\otimes$ Au+Au MB embedding reference  as shaded boxes.

The results show within uncertainties no modification in the Jet Splitting Function as measured via 
SoftDrop for the selected hard core di-jet sample.
This finding is highlighted in the ratios shown in Fig.~\ref{fig:auau}, which supplements 
the data shown in the presentation.
We note that toy model calculations suggest that potential anti-correlated systematics not yet considered in this preliminary analysis
would disproportionally affect the the right-most bin.

\section{Summary}
\label{sec:summary}
We presented the first measurement of $z_g$ in p+p collisions at 200 GeV in
a $p_T^{\text{part}}$ range between 10 and 30 GeV/$c$. After bin-by-bin correction, the distributions 
of the trigger jet are consistent with those of recoil jets not containing an $E_T=5.4$ GeV High Tower.  
The $z_g$ measurements over the entire kinematic range are in good agreement with PYTHIA simulations.

In Au+Au collisions, a set of ``hard core'' di-jets that were previously found to be significantly imbalanced with respect to an embedded p+p reference,
was examined with the added requirement of the High Tower being contained in the trigger jet. 
Within uncertainties, neither trigger nor recoil side $z_g$ measurements displayed modifications compared to the reference. 

In a similar study, the CMS collaboration first reported significant modifications
of the Jet Splitting Function in central Pb+Pb collisions at 5~TeV~\cite{CMS:2016jys}. 
Remarkably, the lowest reported $p_{T,\text{Jet}}$ bin between 140 and 160~GeV/$c$ 
displayed the strongest modification while above ca. 200~GeV/$c$ the ratio between Pb+Pb
and the p+p reference tapered off to unity.

A possible reason that our di-jet selection does not exhibit such a modification
may be that the selection is dominated by unmodified or only mildly modified jets.
Another explanation may arise because $z_g$ approximates the earliest or hardest split in the measured kinematic range,
which may in fact occur mostly outside of the medium. 
If that is the case, a measurement of the groomed soft energy may be directly correlated to in-medium 
gluon radiation.
Additional close collaboration with the theory community will be needed to interpret our findings,
as well as future precision improvements that utilize the recent high-statistics data sets collected by the STAR experiment.

%%\nocite{*}
\bibliographystyle{elsarticle-num}
%\bibliography{jos}
\bibliography{zgref}






% \end{thebibliography}

\end{document}